\documentclass[sigconf]{acmart}
\usepackage{multirow} 
\usepackage{array}
\usepackage{makecell}
\usepackage{subcaption}
\usepackage{enumitem}

\AtBeginDocument{%
  \providecommand\BibTeX{{%
    \normalfont B\kern-0.5em{\scshape i\kern-0.25em b}\kern-0.8em\TeX}}}

\setcopyright{acmcopyright}
\copyrightyear{2018}
\acmYear{2018}
\acmDOI{XXXXXXX.XXXXXXX}

\acmConference[Conference acronym 'XX]{}{June 03--05,
  2018}{Woodstock, NY}
\acmPrice{15.00}
\acmISBN{978-1-4503-XXXX-X/18/06}

\begin{document}


\title{Mixed Supervised Graph Contrastive Learning \\for Recommendation}



\author{Weizhi Zhang}
\email{wzhan42@uic.edu}
\affiliation{%
  \institution{University of Illinois Chicago}
  \city{Chicago}
  \country{USA}}

\author{Liangwei Yang}
\email{lyang84@uic.edu}
\affiliation{%
  \institution{University of Illinois Chicago}
  \city{Chicago}
  \country{USA}}

\author{Zihe Song}
\email{zsong29@uic.edu}
\affiliation{%
  \institution{University of Illinois Chicago}
  \city{Chicago}
  \country{USA}}

\author{Henry Peng Zou}
\email{pzou3@uic.edu}
\affiliation{%
  \institution{University of Illinois Chicago}
  \city{Chicago}
  \country{USA}}

\author{Ke Xu}
\email{kxu25@uic.edu}
\author{Yuanjie Zhu}
\email{yzhu224@uic.edu}
\affiliation{%
  \institution{University of Illinois at Chicago}
  \city{Chicago}
  \country{USA}}

\author{Philip S. Yu}
\email{psyu@uic.edu}
\affiliation{%
  \institution{University of Illinois at Chicago}
  \city{Chicago}
  \country{USA}}


\begin{abstract}

Recommender systems (RecSys) play a vital role in online platforms, offering users personalized suggestions amidst vast information. Graph contrastive learning aims to learn from high-order collaborative filtering signals with unsupervised augmentation on the user-item bipartite graph, which predominantly relies on the multi-task learning framework involving both the pair-wise recommendation loss and the contrastive loss. This decoupled design can cause inconsistent optimization direction from different losses, which leads to longer convergence time and even sub-optimal performance. Besides, the self-supervised contrastive loss falls short in alleviating the data sparsity issue in RecSys as it learns to differentiate users/items from different views without providing extra supervised collaborative filtering signals during augmentations. In this paper, we propose Mixed Supervised Graph Contrastive Learning for Recommendation (MixSGCL) to address these concerns. MixSGCL originally integrates the training of recommendation and unsupervised contrastive losses into a supervised contrastive learning loss to align the two tasks within one optimization direction. To cope with the data sparsity issue, instead unsupervised augmentation, we further propose node-wise and edge-wise mixup to mine more direct supervised collaborative filtering signals based on existing user-item interactions. Extensive experiments on three real-world datasets demonstrate that MixSGCL surpasses state-of-the-art methods, achieving top performance on both accuracy and efficiency. It validates the effectiveness of MixSGCL with our coupled design on supervised graph contrastive learning.

\end{abstract}





\maketitle

\section{Introduction}

Recommender systems (RecSys) have emerged as a cornerstone in the architecture of modern online systems and web applications, playing a pivotal role in filtering and personalizing the information presented to users. These systems are engineered to deliver tailor-made item recommendations that align with individual user preferences. The scope of RecSys extends across various domains, including but not limited to, digital retailing~\cite{wang2020time,hwangbo2018recommendation,zhang2023dual}, social networking platforms~\cite{jamali2010matrix,fan2019graph}, and video-sharing services~\cite{wei2023multi}. Among the methodologies employed in RecSys, collaborative filtering (CF) stands out as a prominent approach~\cite{xu2023graph,yang2022large}. CF algorithms function by extrapolating patterns from historical user-item interactions to identify and recommend items to users with similar interaction profiles. However, this approach is inherently challenged by the issue of data sparsity~\cite{yang2023graph}, a prevalent problem in RecSys. Data sparsity arises due to the limited interactions that each user typically has with the items, thus constraining the system's ability to model and predict user preferences accurately.

Graph contrastive learning has been increasingly integrated into RecSys as a solution to mitigate the data sparsity challenge. This approach primarily involves data augmentation on bipartite graph structure, where high-order structural signals are extracted via graph neural networks~\cite{he2020lightgcn,wang2019neural,ying2018graph}. Concurrently, graph contrastive learning establishes a self-supervised learning task, aimed at enhancing representation expressiveness by distinguishing the nodes in the bipartite graph. This is achieved through the creation of varied varied representation from different views, derived from graph augmentation techniques or embedding perturbations. Existing methodologies in this domain~\cite{wu2021self,lin2022improving,cai2023lightgcl,yu2022graph, yang2023generative} typically employ a decoupled design approach. This involves the utilization of a recommendation-specific supervised loss function~\cite{rendle2012bpr}, designed to learn from user-item interaction signals, in conjunction with a distinct contrastive learning loss~\cite{oord2018representation}, which focuses on assimilating knowledge from self-supervised paradigm. These dual loss functions are simultaneously optimized during the training phase, enabling the model to independently approximate two objects.

Despite their enhanced performance over traditional models that solely consider original user-item interactions~\cite{ying2018graph, wang2019neural, he2020lightgcn}, current graph contrastive learning methods in RecSys still confront two significant challenges that impede further improvements in performance and efficiency.
1) \textit{Inconsistent Gradients}. It arises from the prevalent practice of employing a decoupled design for high-order supervision and self-supervised contrastive signals. This approach necessitates the development of two distinct loss functions. These functions, while aiming to learn from different tasks, optimize the same set of parameters, leading to potential conflicts and inconsistent gradients. Such inconsistent gradients have been identified as a primary factor contributing to unstable training and performance degradation, as highlighted in recent studies~\cite{liu2023deep, tang2023improving}. This not only expends considerable computational resources but also hampers further enhancement and training efficiency of the model.
2) \textit{Unsupervised Augmentation}. It is inherent in these methods under the contrastive learning framework. Typical approaches involve unsupervised manipulations of the graph adjacency matrix \cite{wu2021self, cai2023lightgcl, yang2023generative} or the introduction of noise in node embeddings~\cite{yu2022graph, yang2023generative}. Such unguided and randomized augmentation processes risk significant information loss, as evidenced in scenarios where influential nodes or critical edges are omitted~\cite{wu2021self}. Moreover, in the context of sparse user-item interaction data, unsupervised augmentation can introduce considerable noise, adversely impacting the model's training efficiency. This is further compounded by the need for additional epochs to achieve convergence, owing to the model's reliance on learning from randomly augmented data at each iteration.

In this paper, we introduce Mixed Supervised Graph Contrastive Learning for Recommendation (MixSGCL), a novel framework designed to address the two identified obstacles. To tackle the issue of inconsistent gradients, MixSGCL incorporates a unique supervised graph contrastive learning (SGCL) loss. This loss function integrates the user-item interaction supervision signal directly within the self-supervised loss, thereby merging supervised and self-supervised contrastive learning tasks into a single optimization objective. Such an approach not only mitigates the issue of inconsistent gradients arising from disparate loss components but also obviates the necessity for extensive hyperparameter tuning to balance different loss weights. We further conduct a theoretical analysis in comparison with the SSLRec loss in \cite{wu2021self, lin2022improving, cai2023lightgcl, yang2023generative, yu2022graph}, elucidating the benefits of SGCL from avoiding the user-item distribution shift during training, fostering better and distinguishable representations for ranking. 
Moreover, MixSGCL innovatively enhances supervision signals from the ground-truth user-item interactions to address the data sparsity challenge. Diverging from the unsupervised augmentation, our method leverages mixup \cite{zhang2017mixup}, specifically tailored for RecSys, to build external supervision signals guided by existing user-item interactions. We propose two novel supervised augmentation strategies: node-level and edge-level mixup. Node-level mixup augments node representations from different graph convolution layers, creating diversified representations for the same node. Conversely, edge-level mixup blends node representations along an edge, effectively simulating more user-item interactions. These mixup techniques efficiently harness supervised historical interactions, augmenting more direct and reliable supervision signals compared to unsupervised augmentation. In addition, since SGCL loss computations eliminate the negative sampling process and two types of mixup-based supervised augmentations are conducted after the graph convolution, the overall training is extremely time-efficient.
Our contributions are summarized as follows:

\begin{itemize}[leftmargin=*]
    \item Identification of Key Issues: We have conceptually pinpointed critical challenges in graph contrastive learning for recommendation, specifically the \textit{Inconsistent Gradients} and \textit{Unsupervised Augmentation} issues, which have been impeding the achievement of enhanced performance and efficiency in RecSys.
    \item Development of the novel supervised graph contrastive loss: Methodologically, we have developed and theoretically analyzed an innovative supervised graph contrastive learning loss. This loss uniquely integrates the training of supervised and self-supervised contrastive tasks within a singular objective function, marking a significant methodological advancement in this domain.
    \item Graph-based Mixup: We introduce the node and edge-level mixup on user-item bipartite graphs. This method provides additional direct and reliable supervision signals based on historical user-item interactions to cope with the unsupervised augmentation and data sparsity issue.
    \item Experimentally, we conduct extensive experiments on three real-world datasets to test the effectiveness of MixSGCL. It achieves the highest recommendation accuracy with the lowest training time, demonstrating the remarkable performance of MixSGCL.
\end{itemize}

\section{Preliminaries}

\subsection{Problem Definition}
The problem definition of graph-based recommendation involves harnessing inherent graph structures to enhance the personalization accuracy in RecSys. 
Formally, the problem can be defined as follows.
Consider the bipartite graph structure of user-item interactions $\mathcal{G}=(\mathcal{V}, \mathcal{E})$
where the graph encompasses nodes including users $u_i \in \mathcal{V}_u$ and items $v_j \in \mathcal{V}_v$ and edges representing relationships or interactions between these nodes such as $(u_i, v_j) \in \mathcal{E}$, the objective is to develop a recommendation algorithm that utilizes the graph structure to predict and rank items that a user might be interested in but has not interacted with yet.

\subsection{Self-supervised Graph Recommendation}
Recent approaches \cite{wu2021self,lin2022improving,yu2022graph,cai2023lightgcl, yang2023generative} for self-supervised graph recommendation all rely on the multi-task learning framework and jointly learn from contrastive learning (self-supervised) tasks and supervised recommendation tasks on the node representations.

In a formal definition, the joint learning scheme in self-supervised graph recommendation is to optimize the SSLRec loss function:
\begin{equation}
    \mathcal{L}_{\text {sslrec}}=\mathcal{L}_{\text {rec }}+\lambda \mathcal{L}_{gcl},
\end{equation}
where the $\mathcal{L}_{\text {sslrec}}$ consists of the traditional recommendation loss $\mathcal{L}_{\text {rec }}$, such as BPR \cite{rendle2012bpr}, and the contrastive learning (CL) loss $\mathcal{L}_{gcl}$, with $\lambda$ to weight the two losses. As the inputs are the bipartite user-item interaction graph, the contrastive learning process is conducted concurrently for both the user and item side. Based on the InfoNCE \cite{oord2018representation}, the user-side CL loss $\mathcal{L}^u_{gcl}$ is formulated as:
\begin{equation}\label{eq: sgl}
    \mathcal{L}^u_{gcl}=\sum_{i \in \mathcal{B}_u}-\log \frac{ \exp \left(u_i^{\prime \top} u_i^{\prime \prime} / \tau\right)}{\sum_{k \in \mathcal{B}_u} \exp\left(u_i^{\prime \top} u_k^{\prime \prime} / \tau\right)},
\end{equation}
where $i, k$ are the nodes (users) in the user batch $\mathcal{B}_u$ and $u_i^{\prime}, u_i^{\prime \prime}$ are two views of node representation after the data augmentations. Note that the $\tau$ is a hyperparameter that controls the temperature of the InfoNCE loss and applies the penalties on hard negative samples. Then, the item-side CL loss $\mathcal{L}^v_{gcl}$ can be calculated in the same format. Finally, the CL loss will be the summation of both user and item sides $\mathcal{L}_{gcl} = \mathcal{L}^u_{gcl}+\mathcal{L}^v_{gcl}$.
In general, the graph contrastive learning loss $\mathcal{L}_{gcl}$ encourages the similarity between two variant embeddings $u_i^{\prime}$ and $u_i^{\prime \prime}$ from the same node while maximizing the inconsistency among all the other nodes in batch.

\section{Proposed solutions}

{In this section, we introduce the proposed MixSGCL in detail. Firstly, we demonstrate the novel Supervised Graph Contrastive Loss that integrates the training of both supervised recommendation and contrastive learning losses. We conduct further in-depth theoretical analysis of SGCL and compare it with current SSLRec loss~\cite{wu2021self}. Then we demonstrate the proposed Node/Edge-level Mixup to provide abundant supervision signals to assist the training. Besides, a detailed time complexity investigation is conducted to demonstrate the efficiency of the proposed method.}

\subsection{Supervised Graph Contrastive Learning}
In the context of graph-based recommendation tasks, suppose we observe a pair of interacted users and items with corresponding initial input ID embeddings $u^0_i$ and $v^0_j$, and we adopt the light graph convolution as in \cite{he2020lightgcn}, by consistently aggregating the connected neighbors' representations:
\begin{equation}
\begin{aligned}
u^{k+1}_i = \sum_{j\in N_i} \frac{1}{\sqrt{|N_i|} \sqrt{|N_j|}}v^{k}_j,\\
v^{k+1}_j = \sum_{i\in N_j} \frac{1}{\sqrt{|N_i|} \sqrt{|N_j|}}u^{k}_i,\\
\end{aligned}
\end{equation}
where $u^{k}_i$ and $v^{k}_j$ are embeddings of user $u_i$ and item $v_j$ at $k$-th layer, respectively. The normalization employs the average degree $\frac{1}{\sqrt{|N_i|} \sqrt{|N_j|}}$ to temper the magnitude of popular nodes after graph convolution in each layer.
Afterward, the collaborative filtering final embedding is obtained by synthesizing the layer-wise representations:
\begin{equation}
\begin{aligned}
u_i = \sum^{K}_{k=0} \alpha_k u^k_i; \quad
v_j = \sum^{K}_{k=0} \alpha_k v^k_j,
\end{aligned}
\end{equation}
where $\alpha_k$ is the representation weight of the $k$-th layer and $K$ is total numbers of the layers. Note that we assume $\alpha_k = 1/K$ unless mentioned in the following sections.

For the self-supervised graph recommendation, the contrastive loss $\mathcal{L}_{gcl}$ in Equation~\ref{eq: sgl} is incapable of explicitly learning supervised signals from existing interactions within graph data. Therefore, recommendation loss $\mathcal{L}_{rec}$ is required for label utilization, whilst an extra weight $\lambda$ should be dedicatedly tuned to balance and mitigate the gradient inconsistency. Towards uniting the power of the supervised signal and self-supervised learning strategy in the graph recommendations, we devised a new type of supervised graph contrastive learning (SGCL) loss as follows:
\begin{equation}
\mathcal{L}_{sgcl}=\sum_{(i, j) \in \mathcal{B}}-\log \frac{\exp \left({u}_i^{\top} {v}_j / \tau\right)}{
\sum_{(i^{\prime},j^{\prime}) \in \mathcal{B}} \left(\exp ( {u}_i^{\top}  {u}_{i^{\prime}}  / \tau) + \exp ( {v}_j^{\top}  {v}_{j^{\prime}}  / \tau)\right)}, 
\end{equation}
where $(i, j)$ are paired user-item interaction in batch $\mathcal{B}$ and $(i^\prime, j^\prime)$ are the rest user-item representations in batch. It is noted that this version of the loss function does not involve any type of data augmentation, and the user-item interactions are considered positive pairs. Therefore, both complex graph augmentation and time-consuming negative sampling processes are avoided, making the SGCL training process extremely efficient. More importantly, we alleviate the need to use multi-task learning for model optimization and use only one supervised graph contrastive loss during training.

\begin{figure*}[htpb]
    \centering
    \includegraphics[width=0.85\linewidth]{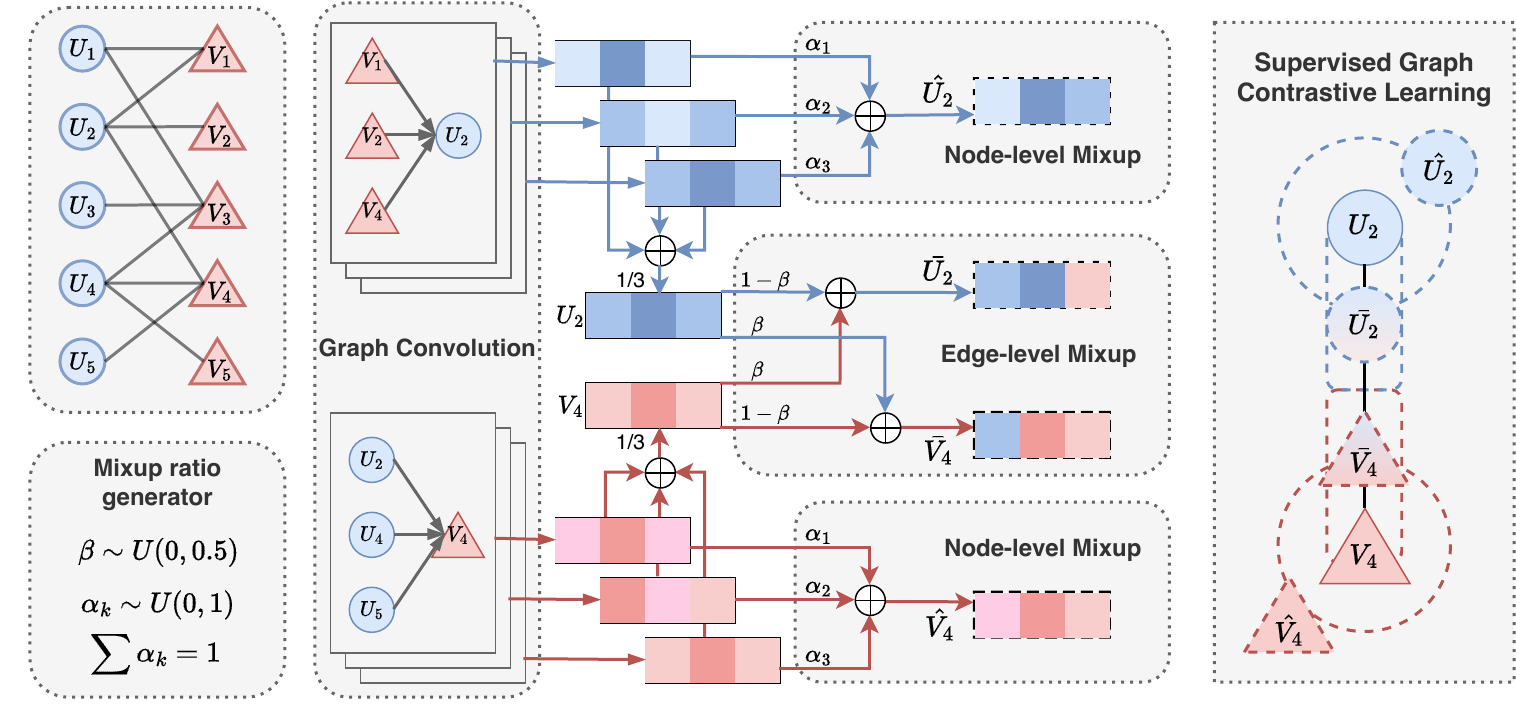}
    \caption{The overall architecture of MixSGCL, including SGCL loss along with the Node/Edge-level Mixup for supervised augmentation. The pair of $(u_2,v_4)$ are augmented for each type mixup once, generating two extra pairs $(\hat{u}_2, \hat{v}_4)$ and $(\bar{u}_2, \bar{v}_4)$ supervision signals for SGCL loss optimization. The rightmost figure demonstrates their relations in the embedding space.}
    \label{fig: mixsgcl}
\end{figure*}

\subsection{Analysis of SGCL and SSLRec} \label{sec: theory}

Assume given a pair of user-item $(u, v)$ and a sampled negative item $v^{-}$ and the corresponding final embeddings are normalized; minimizing the SSLRec loss without augmentation, we have: 
\begin{equation}
\label{eq: ssl_loss}
\begin{aligned}
& \mathcal{L}_{sslrec} = -\log \left(\frac{1}{1 +\exp({u^\top v^-} - {u^\top v)}}\right) - \lambda \log \frac{ \exp \left(u^ \top u / \tau\right)}{\sum_{u^\prime \in \mathcal{B}_u} \exp\left(u^\top u^\prime / \tau\right)} \\
& \hspace{0.5cm} - \lambda \log \frac{ \exp({v^{\top} v / \tau})}{\sum_{v^\prime \in \mathcal{B}_v} \exp({v^{\top}  v^{\prime} / \tau})} \\
& = \log \left(\exp({u^T v})+ \exp({u^T v^-})\right)  -u^T v
+ \lambda \log {\sum_{u^\prime \in \mathcal{B}_u} \exp\left(u^\top u^\prime / \tau\right)} \\
& \hspace{0.5cm}+ \lambda \log {\sum_{v^\prime \in \mathcal{B}_v} \exp({v^{\top}  v^{\prime} / \tau})}  - \frac{2\lambda}{\tau} \\
& \geq \log \left(\exp({1})+ \exp({u^T v^-})\right) + \lambda \log {\sum_{u^\prime \in \mathcal{B}_u} \exp\left(u^\top u^\prime / \tau\right)}   \\
& \hspace{0.5cm} + \lambda \log {\sum_{v^\prime \in \mathcal{B}_v} \exp({v^{\top}  v^{\prime} / \tau})}  - \frac{\tau+2\lambda}{\tau},
\end{aligned}
\end{equation}
where the equality holds when the model is fully trained with perfect supervision and obtains $u^{T}v = 1$. Then we get the final form and optimization object of the SSLRec loss as minimizing the $u^Tu^\prime$, $v^Tv^\prime$, and $u^Tv^{-}$. These essentially correspond to distancing the embeddings among users, items, and users-items, respectively. 
On the one hand, both users and items become uniformly distributed in the entire feature space. On the other hand, as only sparse interactions are observed, distancing user-item pairs that lack observed interactions leads to similar effects of pushing the embedding spaces of users and items away, resulting in the distribution shift.
The inconsistent gradients problem emerges as one encourages the users and items to co-exist sparsely on the entire embedding space while the other opposes the overlapping feature space of users and items.

Whereas, given the same condition, SGCL is be simplified as:
\begin{equation}
  \mathcal{L}_{sgcl} \geq \log\left( {\sum_{(u^{\prime}, v^{\prime})  \in \mathcal{B}} \exp{\left( u^{\top}  u^{\prime} / \tau\right)}}   
 + \exp{ \left( v^{\top} v^{\prime} / \tau\right)} \right) -1/\tau.
\end{equation}
Considering the final form of our SGCL loss, instead of continuously separating the user and item embedding spaces away, SGCL only differentiates the user from the users' embedding space and the item from the items' embedding space, respectively. This contributes to a uniform distribution of users and items to facilitate the ranking in the evaluation, whilst keeping the users' and items' representations in the same embedding space with alignment forced by the supervision signals.

\subsection{Mixup for Supervised Augmentation}
In the self-supervised graph recommendation, contrastive learning is often applied with different sets of data augmentations. Unfortunately, none of them attempt to augment graph data under the existing supervised label nor by adding the supervised signals (more interactions) to guarantee \textbf{abundant high-quality supervision} toward the encoder training. Motivated by the mixup \cite{zhang2017mixup} technique, we aim to create new and virtual interactions samples. 
However, the inter-class mixup in computer vision is infeasible to be directly applied for recommendation, and instead we propose the intra-class mixup: only mixing the samples of the positive classes from node-level or edge-level for supervised augmentation on supervision signals.
Virtual users/items can be created by mixing the layer-wise representations of themselves or by two adjacent nodes, thereby creating a new edge for training while not increasing the node embeddings for training. In Figure ~\ref{fig: mixsgcl}, we demonstrate an example of how to create those virtual nodes via mixup-based augmentation. For the pair of $(u_2,v_4)$, two types of augmentations are conducted once to generate two virtual interactions. For a batch $\mathcal{B}$ data, $N_{mix}$ times of augmentations can be implemented to produce $2N_{mix} \mathcal{B}$ of new data during SGCL optimization. Note that in the following sections, we denote the \textbf{SGCL} as the model purely relying on the supervised graph contrastive learning loss $\mathcal{L}_{sgcl}$, and the \textbf{MixSGCL} as the version armed with supervised augmentations.

\subsubsection{\textbf{Node-level Mixup (NMix)}} In the SGCL paradigm, the supervision signal is regarded as existing user-item interactions. Thus, augmenting supervision revolves around the creation of new user-item pairs. Considering a known user-item interaction $(u_i, v_j)$, the straightforward way is to craft a virtual user and item akin to $u_i$ and $v_j$.
To this end, we propose the node-level mixup for augmentation - for a batch $\mathcal{B}$, each user $u_i$ will be reconstituted from their layer representation to produce a virtual counterpart user $\hat{u}_i$ of the original user $u_i$. Notably, the layer weight $\alpha_k$ is reinterpreted as a node-wise mixup ratio, sampled from a uniform distribution:
\begin{equation}
\begin{aligned}
 \alpha_k \sim U(0, 1); \quad \sum^{K}_{k=0} \alpha_k  = 1.
\end{aligned}
\end{equation}
The generation of virtual nodes is achieved by remixing the layer-wise representation. By assigning differential weight to different hops' receptive fields in graph convolution, more structural information is gathered during the learning phase. Furthermore, as illustrated in Figure~\ref{fig: mixsgcl}, the node-level mixup can be conceptualized as generating virtual nodes $\hat{U}_2$ and $\hat{V}_4$ surrounding the ${U}_2$ and ${V}_4$. This proximity facilitates a smoother and expansively explored embedding space around those original nodes that are refined and aligned to be closer.

\subsubsection{\textbf{Edge-level Mixup (EMix)}}
Beyond the node-level mixup of the layer-wise representations, there remains an underutilization of the existing pair-wise representations that characterize observed user-item interactions. As the goal of the recommendation is to encourage the node representations along an edge to be similar, we propose to explicitly add more supervision directly along the edge. Treating $u_i$ and $v_j$ as two endpoints, new interactive connections overlapping with the original edges are created for each batch of data. Specifically, for a batch $\mathcal{B}$, each user-item pair $(u_i, v_j)$ can be amalgamated to synthesize two corresponding virtual user/item nodes on the edge:
\begin{equation}
\begin{aligned}
\bar{u}_i = (1-\beta) u_i + \beta v_j,\\
\bar{v}_i = \beta u_i + (1-\beta) v_j,\\
\end{aligned}
\end{equation}
where $\beta$ is a randomly generated ratio number drawn from the uniform distribution 0 to 0.5, $\beta \sim U(0, 0.5)$. It is noteworthy that $\bar{u}_i$ can be interpreted as the user with stronger favors of the products similar to $v_j$, and the  $\bar{v}_j$ is the items adding more characteristics appealing to the type user $u_i$. As depicted in Figure~\ref{fig: mixsgcl}, a virtual edge is established by connecting $\bar{U}_2$ and $\bar{V}_4$, imposing more supervised signals on the edge delineated by the user-item pair.


\subsection{Time Complexity}
In this subsection, we conduct the time complexity analysis on SGCL and MixSGCL and compare them with one of the predominant baseline methods SGL \cite{wu2021self} in the self-supervised graph recommendation. We first denote $|\mathcal{E}|$ as the number of edges in the graph. Then, let K be the number of layers in graph convolution, and d be the embedding size. $p$ represents the probability of retaining the edges in SGL and $N_{mix}$ indicates the times of augmentations applied in MixSGCL.

Then, we can derive that: 1) For the normalization of the adjacency matrix, both SGCL and MixSGCL do not contain the graph structure manipulation; they only need to compute the $2|\mathcal{E}|$ non-zero elements in the original adjacency matrix. On the contrary, SGL drops the edges in the graph data to create two versions of the modified adjacency matrix, and thus the time complexity is approximately three times of our methods. 
2) In the graph convolution, as SGL changed the graph structure twice in contrastive learning, it requests an extra $4p|\mathcal{E}|Kd$ for graph convolution. Whereas no augmentations are implemented before graph convolution in SGCL and MixSGCL, and they only convolution once based on the LighGCN backbone, resulting in efficient $2|\mathcal{E}|Kd$ time consumption. 
3) Both SGCL and MixSGCL obviate the necessity of BPR loss computation, while SGL still needs $2Bd$ for calculating the 2B pairs of positive and negative interactions.
4) Regarding the graph contrastive loss, SGL separates the user and item loss calculation into $\mathcal{L}^u_{gcl}$ and $\mathcal{L}^v_{gcl}$, with $Bd$ for the numerator part and $B^2d$ for the denominator part computation, and the total time cost is $O(2Bd + 2B^2d)$. Since SGCL is based on a joint embedding learning process of the users and items, the time complexity in the batch is $O(Bd + 2B^2d)$. On account of the number of mixup-based augmentations, MixSGCL will cost $O(N_{mix}Bd + 2N_{mix}B^2d)$ time per batch. It is noted that, as all experiments are conducted on the GPU parallel computation, therefore the time spent in loss computation differs slightly.

\begin{table}[ht]
\centering
\caption{Time complexity comparison of SGL, SGCL, and MixSGCL in different steps of graph recommendation.}

\begin{tabular}{l|m{2.2cm}|m{1.8cm}|m{1.8cm}}
\toprule
 Steps & SGL & SGCL & MixSGCL\\
\midrule
\makecell[l]{Adj.\\Matrix} & \( O((2 + 4p)|\mathcal{E}|) \) & \( O(2|\mathcal{E}|) \) & \( O(2|\mathcal{E}|) \)\\
\midrule
\makecell[l]{Graph \\ Conv.} & \( O((2 + 4p)|\mathcal{E}|Kd) \) & \( O(2|\mathcal{E}|Kd) \)  & \( O(2|\mathcal{E}|Kd) \)\\
\midrule
BPR Loss  & \( O(2Bd) \) & - & -\\
\midrule
GCL Loss  & \( O(2Bd + 2B^2d) \) & \( O(Bd + 2B^2d) \) & \( O(N_{mix}Bd + 2N_{mix}B^2d) \) \\
\bottomrule
\end{tabular}
\label{tab:complexity}
\end{table}

\section{Evaluation}

\subsection{Experimental Setup}
\begin{table}
\begin{center}
\caption{The statistics of the datasets.}
  \label{tab: data}
  \begin{tabular}{cclll}
    \toprule
    Dataset & Users & Items & Interactions & Sparsity\\
    \midrule
    Beauty & 22,364 & 12,102 & 198,502 &	99.9267\%\\
    Toys-and-Games &  19,413 & 11,925 & 167,597 & 99.9276\%\\
    Yelp & 77,278 & 45,639 & 2,103,896 & 99.9403\%\\
  \bottomrule
\end{tabular}
\end{center}
\end{table}

\subsubsection{\textbf{Datasets.}}
To evaluate the superior performance and efficiency of the proposed MixSGCL along with SGCL, we conduct the experiments on three public real-world datasets: Amazon-Beauty (Beauty), Amazon-Toys-and-Games (Toys-and-Games), and Yelp-2018 (Yelp), varying in domains and scale.
Beauty and Toys-and-Games are collected from real-world data in Amazon~\footnote{\url{https://jmcauley.ucsd.edu/data/amazon/links.html}} datasets. We use the 5-core filtering setting by eliminating the users/items with lower than five interactions so as to ensure the data quality for testing. 
The Yelp-2018~\footnote{\url{https://www.yelp.com/dataset}} dataset was adopted from the 2018 edition of the Yelp challenge. Where local businesses such as restaurants or bars are viewed as items. We use the 10-core setting for the Yelp-2018 datasets. 
We split all datasets into training, validation, and testing with the ratio (8:1:1), and the statistical information of the three datasets after the preprocessing is summarized in Table ~\ref{tab: data}.

\subsubsection{\textbf{Evaluation Metrics}}
For the evaluation metrics, Recall@K and NDCG@K are adopted for a fair comparison of all the baselines in the top-K recommendation task. K is set as 20 and 50 in the main performance evaluation and is set to 20 by default in the other experiments. The full-ranking strategy \cite{zhao2020revisiting} is adopted for all the experimental studies, i.e., all the candidate items not interacted with the user will be ranked in testing.

\subsubsection{\textbf{Implementation Details}}
We implement our proposed SGCL and MixSGCL and all the baseline methods based on the RecBole \cite{zhao2022recbole2}. For the training of all the baselines, we carefully search their hyperparameters for different datasets to ensure a fair comparison. The batch size and the embedding size are set to 1,024 and 64, respectively. All the models are optimized with Adam optimizer \cite{kingma2014adam} with the learning rate searching in [1e-2, 5e-3, 1e-3, 5e-4, 1e-4]. To prevent overfitting, we adopt the early stop strategy with patience for consistent performance degradation of NGCG@K for 10 epochs. For our methods SGCL and MixSGCL, we tune the temperature $\tau$ in the range of [0.1, 1] and set $N_{mix}$ as 1 for the mixup-based augmentation. In the efficiency analysis, we run one model with a single GPU at a time for fair comparison.

\begin{table*}[htpb]
\begin{center}
\caption{Performance comparison on three datasets in terms of NDCG and Recall.}
  \label{tab:main}
  \begin{tabular}{ccccccccccccc}
    \toprule
    \multirow{2}{*}{Method} & \multicolumn{4}{c}{Beauty} & \multicolumn{4}{c}{Toys-and-Games}  & \multicolumn{4}{c}{Yelp}\\
    
    \cmidrule(r){2-5} \cmidrule(r){6-9} \cmidrule(r){10-13} 
    & R@20 & N@20 & R@50 & N@50 
    & R@20 & N@20 & R@50 & N@50 
    & R@20 & N@20 & R@50 & N@50 \\
    
    \midrule
    BiasMF & 0.1152 & 0.0544 & 0.1754 & 0.0668 & 0.0981 & 0.0479 & 0.1624 & 0.0625 &
    0.0761 & 0.0366 & 0.1216 & 0.0448 \\
    NeuMF & 0.072 & 0.032 & 0.1317 & 0.049 & 0.069 & 0.0345 & 0.1094 & 0.0428 &
    0.0547 & 0.0252 & 0.1157 & 0.0409\\
    \midrule
    
    NGCF & 0.1126 & 0.0517 & 0.1699 & 0.0625 & 0.0944 & 0.0429 &  0.1532 & 0.0561 &
    0.055 & 0.024 & 0.1374 & 0.0497\\
    LightGCN & 0.1224  & 0.0583 & 0.1903 & 0.0718 & 0.1159 & 0.0565 & 0.1759 & 0.068 &
    0.0953 & 0.0472 & 0.1711 & 0.0645\\

    \midrule
    NCL & 0.1264 & 0.0613 & 0.1948 & 0.0721 & 0.117 & 0.0566 & 0.1782 & 0.0686
    & {0.105} & {0.0518} & {0.1839} & {0.0703} \\
    SGL & {0.1335} & {0.0634} & {0.2048} & {0.078} & \underline{0.1248} & \underline{0.0595} & \underline{0.1869} & \underline{0.072} 
    & 0.1021 & 0.0505  & 0.1805 & 0.0688\\
    LightGCL & 0.1278 & 0.0596 & 0.1905 & 0.0697 & 0.1177 & 0.0549 & 0.1772 & 0.0647
    & 0.0949 & 0.0462 & 0.1731 & 0.0643 \\
    SimGCL & \underline{0.1345}	& \underline{0.064} & \underline{0.2058} & \underline{0.0799} & 0.1162 & 0.0593 & 0.1738 & 0.0713 
    & \underline{0.1058} & \underline{0.0527} & \underline{0.1847} & \underline{0.0707}\\
    \midrule
    
    MixSGCL & \textbf{0.143} & \textbf{0.0694} &  \textbf{0.2116} & \textbf{0.084} & \textbf{0.1355} & \textbf{0.0651} & \textbf{0.1972} & \textbf{0.0779} 
    & \textbf{0.1096} & \textbf{0.0541} & \textbf{0.1935} & \textbf{0.0733} \\
  \bottomrule
\end{tabular}
\end{center}
\end{table*}

\subsubsection{\textbf{Baselines}}
We compare our proposed SGCL and MixSGCL with the following various state-of-the-art methods.

\begin{itemize}[leftmargin=15pt]
    \item \textbf{BiasMF} \cite{koren2009matrix} is a typical matrix factorization method that incorporates the bias vectors for both users and items to improve the learning for user preference and item profiles.
    \item \textbf{NeuMF} \cite{he2017neural} presents a novel collaborative filtering method based on neural networks, utilizing a multi-layer perceptron to learn user-item interaction patterns for the final predictions.
    \item \textbf{NGCF} \cite{wang2019neural} introduces a GCN-based recommendation framework that integrates user-item graph structure into collaborative filtering, leveraging high-order connectivity in graphs. 
    \item \textbf{LightGCN} \cite{he2020lightgcn} simplifies the graph convolutional network recommendation, focusing on the linear and efficient propagation of user and item embeddings through the graph structure.
    \item \textbf{NCL} \cite{lin2022improving} proposes to integrate contrastive learning with enriched neighborhood information from the structural and semantic neighbors in the graph-based recommendations.
    \item \textbf{SGL} \cite{wu2021self} presents a self-supervised learning framework for graph recommendation, by leveraging auxiliary CL tasks to capture intrinsic user-item interaction patterns within the graph.
    \item \textbf{LightGCL} \cite{cai2023lightgcl} introduces an efficient graph contrastive learning framework by exclusively utilizing singular value decomposition for adjacency matrix augmentation.
    \item \textbf{SimGCL} \cite{yu2022graph} challenges the necessity of complex graph augmentation and proposes simple graph contrastive learning and noise-based augmentation for graph recommendation.

\end{itemize}

\subsection{Overall Performance Comparison}

In this comprehensive experimental study, we evaluated the performance of several state-of-the-art recommendation methods across three diverse datasets: Beauty, Toys-and-Games, and Yelp, using critical metrics, including the Recall@20, Recall@50, NDCG20, and NDCG@50. Here, we summarize the main observations:

\begin{itemize}[leftmargin=18pt]
\item Predominantly, MixSGCL achieved the highest NDCG and Recall scores across all datasets, highlighting its superior efficacy in recommendation tasks. This outcome not only emphasizes the significance of the proposed mixup augmentation but also validates the efficiency of the supervised graph contrastive learning framework.
\item Among all traditional collaborative filtering baselines, encompassing MF-based and graph-based models, LightGCN emerges as the best competitor in all three datasets, which demonstrates the effectiveness of the light graph convolution \cite{he2020lightgcn}.
\item Most of the self-supervised graph recommender systems consistently outperform the traditional ones. This suggests that the auxiliary contrastive learning task leverages extra graph structure information, boosting the performance of the predictions on unobserved user-item interactions. Among them, SGL is generally the best model, albeit its duplicate graph convolution due to the dropout on the graph structures.
\end{itemize}

\subsection{Ablation Study}
\begin{figure}[htpb]
    \centering
    \includegraphics[width=1\linewidth]{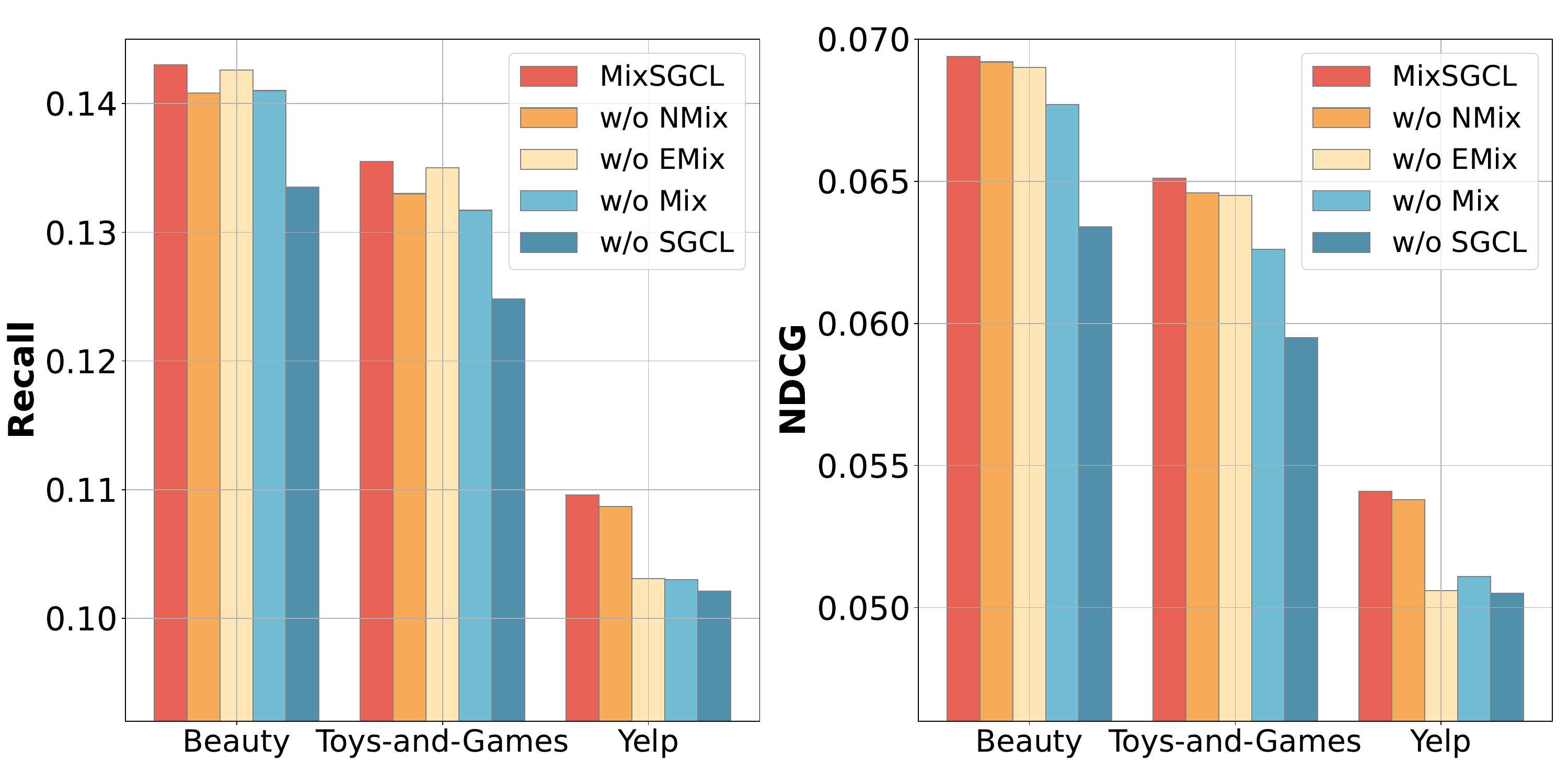}
    \caption{Ablation study on different components.}
    \label{fig: ablation}
\end{figure}

Here, we provide the ablation study of different components of the MixSGCL. From Figure~\ref{fig: ablation}, we observe that MixSGCL achieves the highest recall and NDCG scores on all three datasets, whereas the graph model with SSLRec loss (w/o SGCL) yields extremely low performance. 
Replacing SSLRec loss with our SGCL objective function notably advances performance, even without supervised augmentation. Additionally, Both node-level and edge-level mixup augmentations further bolster the GNN recommendation system, especially on the large and sparse Yelp dataset. Each component within our MixSGCL framework contributes to the final recommendation performance.

\subsection{Efficiency Analysis}
\subsubsection{\textbf{Trade-off between the Performance and the efficiency.}}
\begin{figure}[htpb]
    \centering
    \includegraphics[width=0.85\linewidth]{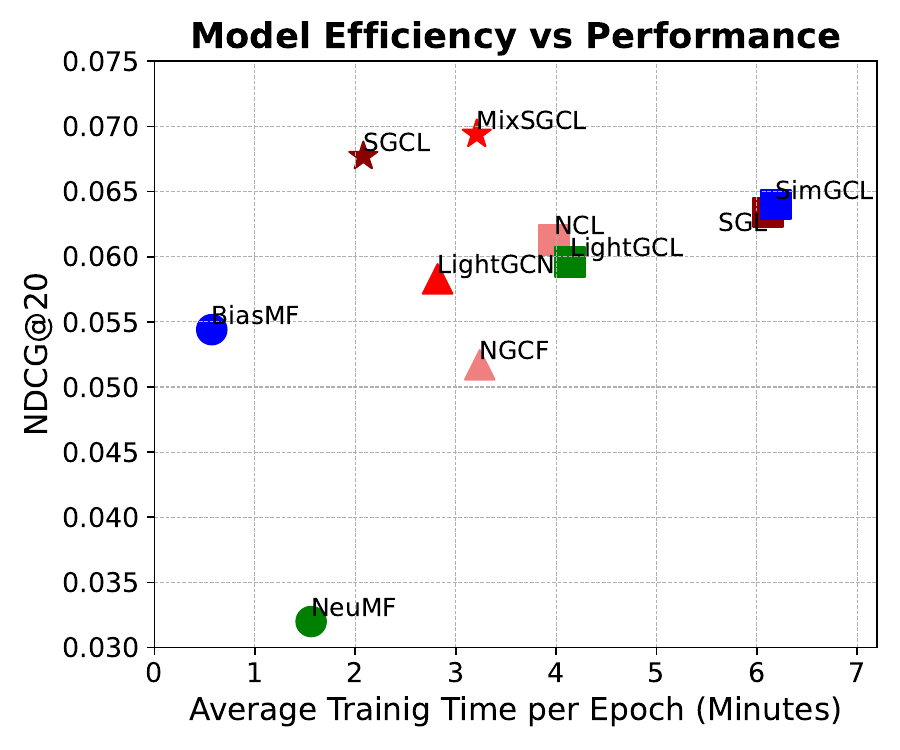}
    \caption{Trade-off between the performance and the efficiency on the Beauty dataset. The upper side indicates better performance; the left side represents more efficient training.}
    \label{fig: tradeoff}
\end{figure}
Figure \ref{fig: tradeoff} demonstrates the superiority of our models in both efficiency and performance. Two matrix factorization models are more efficient than graph-based models but achieve lower performance. Generally, LightGCN and NGCF outperform the rest of the graph contrastive learning baseline frameworks in terms of speed. Our methods exhibit the most competitive performances while being much faster than the other graph contrastive recommendation baselines. With the integration of our supervised augmentation process, the increase in training time is trivial. Our methods maintain a good balance of rapid training efficiency and elevated performance.

\subsubsection{\textbf{Training Time Comparison.}}
In Table~\ref{tab: time}, LightGCN is efficient per epoch but may not always be the best choice when considering total training time due to the higher number of epochs required. SGL appears to offer a better compromise between the number of epochs and time per epoch, often resulting in an approximate total time for the Beauty and Toys-and-Games datasets compared with LightGCN. 
In contrast, SGCL demonstrates extreme efficiency in all datasets, which run for the fewest epochs despite being the fastest per epoch in most cases, thereby resulting in the lowest total training time in all three datasets. Note that, in the Beauty and Yelp datasets, the average training time is lower than LightGCN's due to the presence of a negative sampling process for each epoch for the LightGCN.
MixSGCL, while not excelling in per-epoch efficiency, provides a balanced approach that can lead to a low overall training time, particularly approximating the SGCL in the Beauty dataset.

\begin{table}[htpb] 
\begin{center}
\caption{Training time efficiency comparison on Beauty, Toys-and-Games, and Yelp datasets, which includes the average training time for each epoch, the total number of training epochs, and the total time spent during training (we denote second as s, minute as m, and hour as h as the abbreviation).}
\label{tab: time}
  \begin{tabular}{ccccc}
    \toprule

Dataset  & Method  & Time/Epoch & \# Epochs & Total Time \\
\midrule
\multirow{4}{*}{Beauty} 
&LightGCN    &2.82s	&153&	7.19m \\
&SGL         &6.11s	&63&	6.42m \\
&SGCL        &2.08s	&62&	2.15m \\
&MixSGCL     &3.21s	&52&	2.78m \\
\midrule
\multirow{4}{*}{\makecell[l]{Toys-and\\-Games}} 
&LightGCN    &1.53s	&173&	4.41m \\
&SGL         &6.24s	&75&	7.80m \\
&SGCL        &1.63s	&35&	0.95m \\
&MixSGCL     &2.21s	&50&	1.84m \\
\midrule
\multirow{4}{*}{Yelp} 
&LightGCN    &246.14s	&145&	9.91h \\ 
&SGL         &651.67s	&93&	16.83h \\
&SGCL        &222.07s	&63&	3.89h \\
&MixSGCL     &360.53s	&69&	6.91h \\

  \bottomrule
\end{tabular}
\end{center}
\end{table}

\subsubsection{\textbf{Performance Curve and Convergence Speed}}

\begin{figure}[htpb]
    \centering
    \begin{subfigure}[b]{.95\linewidth}
        \centering
        \includegraphics[width=\textwidth]{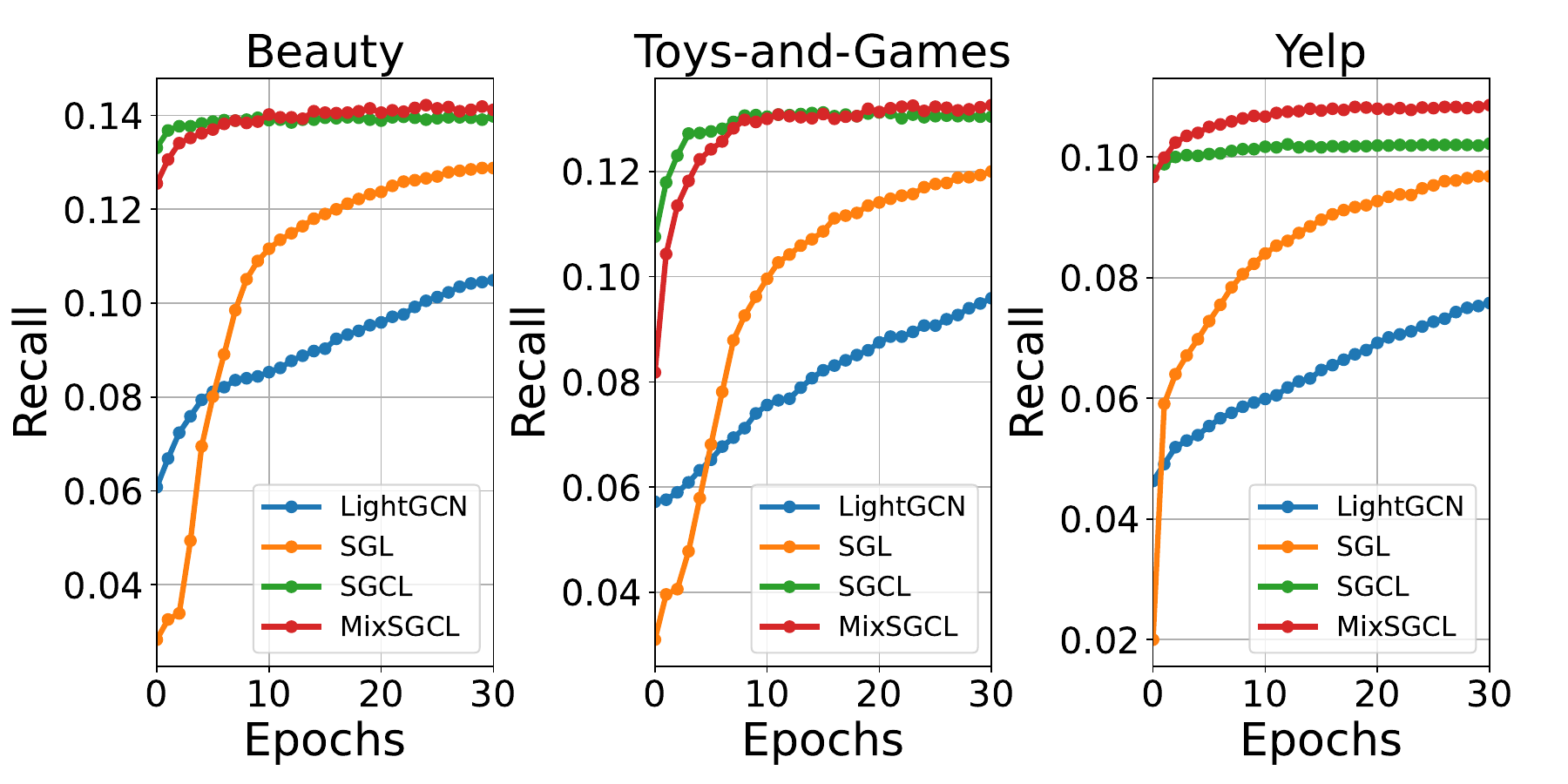} %
    \end{subfigure}
    \hfill 
    \begin{subfigure}[b]{0.95\linewidth}
        \centering
        \includegraphics[width=\textwidth]{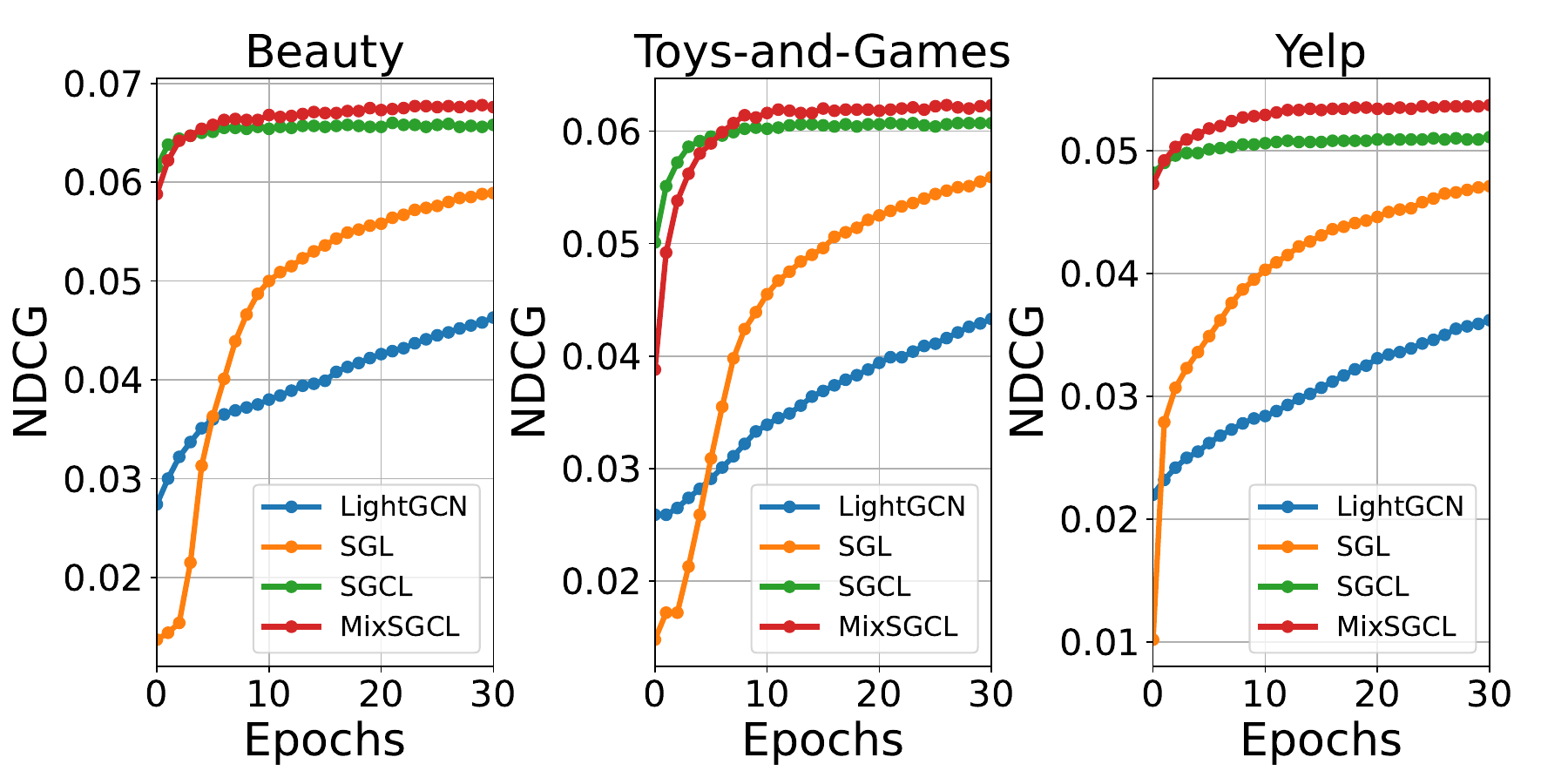} 
    \end{subfigure}
    \caption{Performance curve of the Recall and NDCG in the first 30 epochs.}
    \label{fig: curve}
\end{figure}

We compare the metrics of Recall@20 and NDCG@20 (Figure \ref{fig: curve}) across the first 30 training epochs with the LightGCN and the SGL on three datasets. Our proposed methods reach high recall results at early training stages, SGCL in particular, exhibiting the fastest convergence speed. By 30 epochs, both of our models achieve high and stable Recall levels, whereas the LightGCN and the SGL have not yet arrived at their convergence points. A similar result is observed in the NDCG result figures, where our methods consistently outperform the two baseline models, meanwhile, gaining more performance with mixup augmentations. These findings are in consonance with the outcomes of our training time experiments, underscoring the preeminence of our methodologies in practical and real-world applications.

\subsection{Performance on Data Sparse Scenarios}
To investigate the effectiveness of MixSGCL in the data sparse conditions, we remove different ratios of interactions in the original training data to the testing. In this case, the NDCG might increase if the model maintains robust performance on sparse scenarios, given that more positive items can be ranked in higher positions.
MixSGCL is the most consistent top performer across both metrics in two datasets, with a slight increase in NDCG when the data ratio drops to 75\%, suggesting that it is the most effective method for utilizing sparse amounts of data.
SGL is competitive at full data availability, often close to MixSGCL's performance. However, its performance drops sharply with the decrease of supervision data, surpassed by LightGCN in the 75\% scenario in terms of the Recall and in the 50\% scenario for Recall and NDCG.
This indicates the complex graph augmentation is not only incapable of utilizing the data but also abandons some significant supervision information during unsupervised and randomized augmentation, thus hindering the ranking performance. LightGCN seems to be less sensitive to the decrease in data ratio, with smaller performance degradation, albeit with the lowest performance on original training data.

\begin{figure}[htpb]
    \centering
    \begin{subfigure}[b]{0.9\linewidth}
        \centering
        \includegraphics[width=\textwidth]{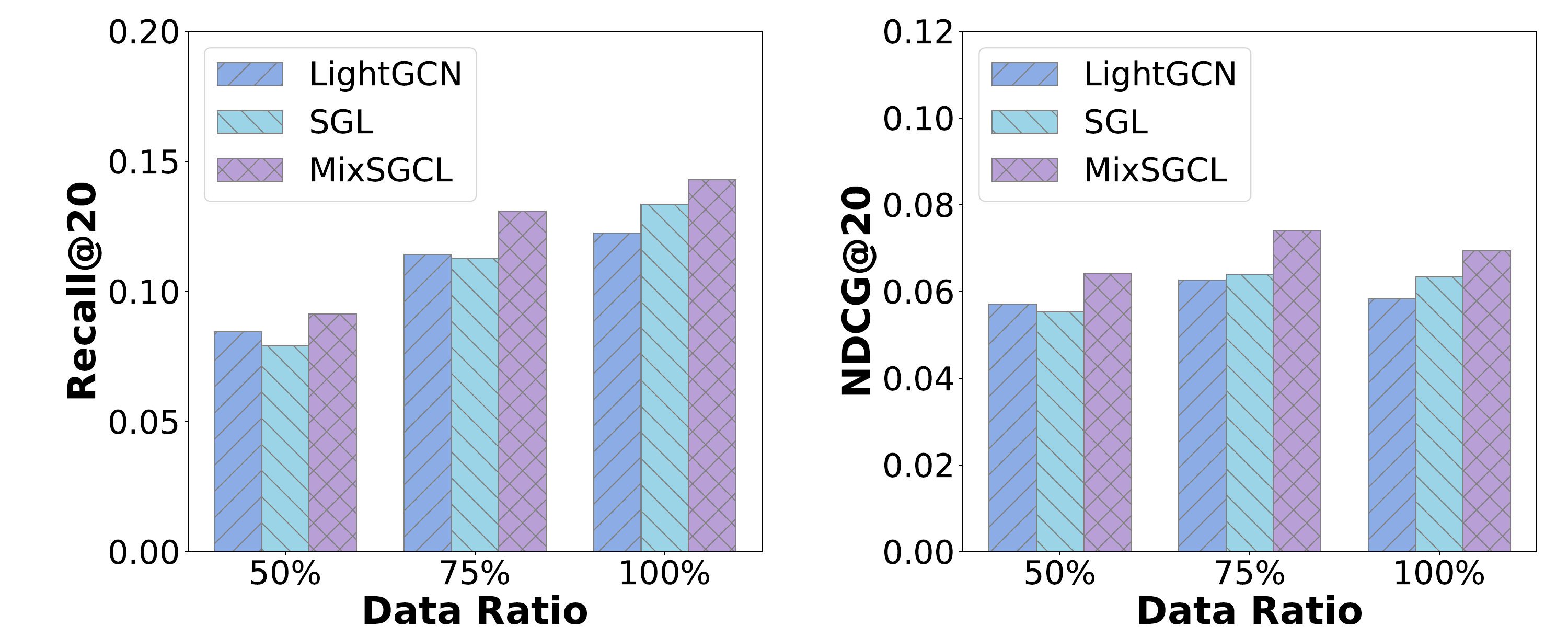} 
        \caption{Different ratios of training interactions in Beauty.}
        \label{fig:sparse_beauty}
    \end{subfigure}
    \hfill 
    \begin{subfigure}[b]{.95\linewidth}
        \centering
        \includegraphics[width=\textwidth]{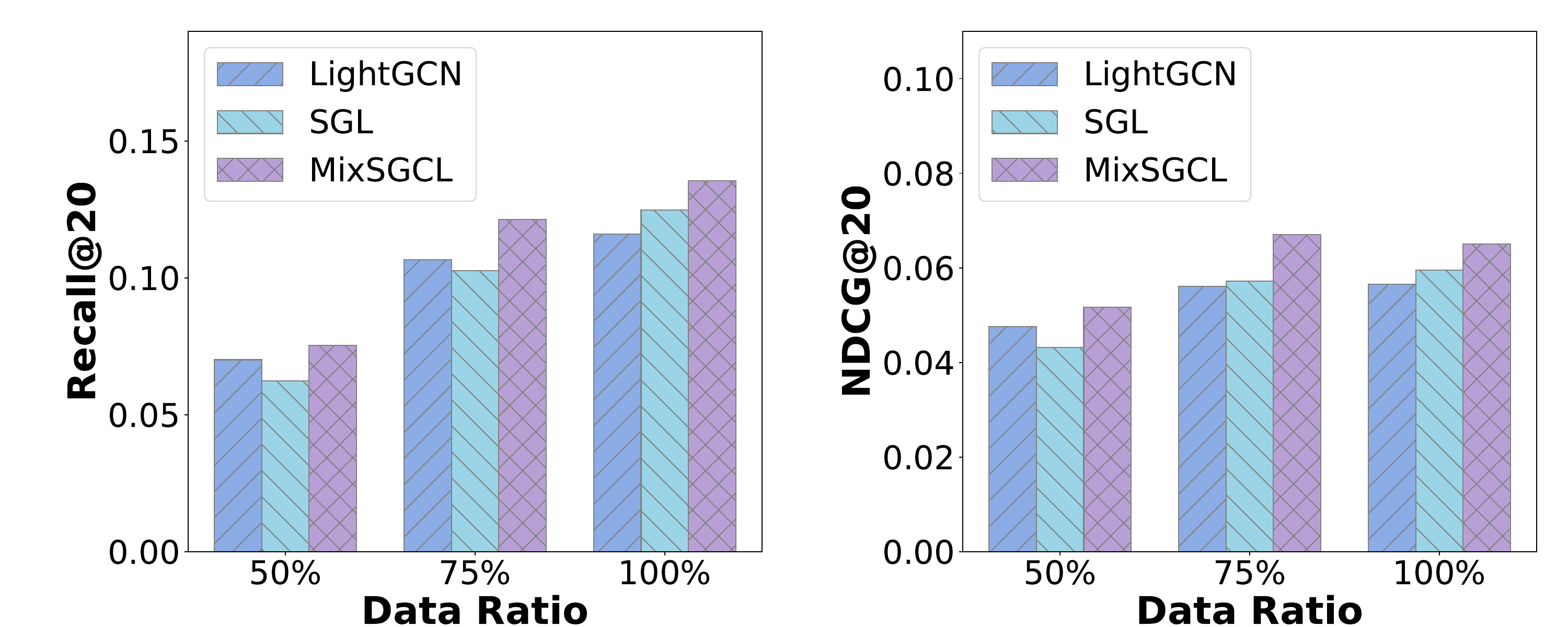} 
        \caption{Different ratios of training interactions in Toys-and-Games.}
        \label{fig:sparse_toys}
    \end{subfigure}
    \caption{Performance trained with different portions of original training data (with more sparse supervision data).}
\end{figure}

\subsection{Visualization of the Embeddings}
In this subsection, we visualize the final user-item embeddings on Beauty and Toys-and-Games datasets via the t-SNE \cite{Poli2019tsne}.
LightGCN exhibits relatively scattered clusters of users and items in the embedding space, with users seemingly more dispersed than items. In the Beauty dataset, a conspicuous cluster of users (circled in red), stands apart from the main distribution, and in the Toys-and-Games dataset, a similar pattern is observed, while the distinct subgroup is more separated from the majority. This pattern indicates that a potential subgroup of users with distinct preferences or items with unique features may not find the corresponding products or customer segments, leading to suboptimal performance outcomes.
The embeddings from the SGL are characterized by a more uniform distribution, with densely populated central clusters, particularly in the Toys-and-Games dataset. Such centralization could signify the model's capability to identify commonality in the popular user-item groups, but the major clusters are not aligned for the user-item sides. In such scenarios, it is hard to distinguish what items are favored by highly active users and what users would prefer to use among all popular items. In comparison, aligned with the theoretical findings in Section~\ref{sec: theory}, the user-item embedding spaces of our methods are highly cohesive and overlap with each other, where the clusters of users and items are perfectly matched cohesively. This indicates much less distribution shift between users and item representations in SGCL and MixSGCL, contributing to better recommendation performances.

\begin{figure*}[ht]
    \centering
    \begin{subfigure}[b]{1.0\linewidth}
        \includegraphics[width=\linewidth]{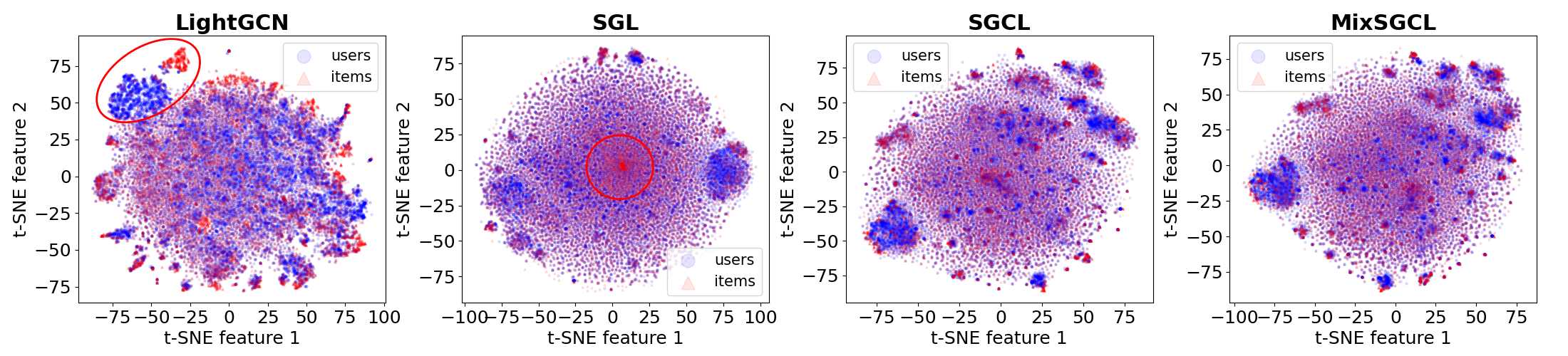}
        \caption{Visualization of the User-Item Embeddings on Beauty Dataset}
        \label{fig: visual_beauty}
    \end{subfigure}
    \begin{subfigure}[b]{1.0\linewidth}
        \includegraphics[width=\linewidth]{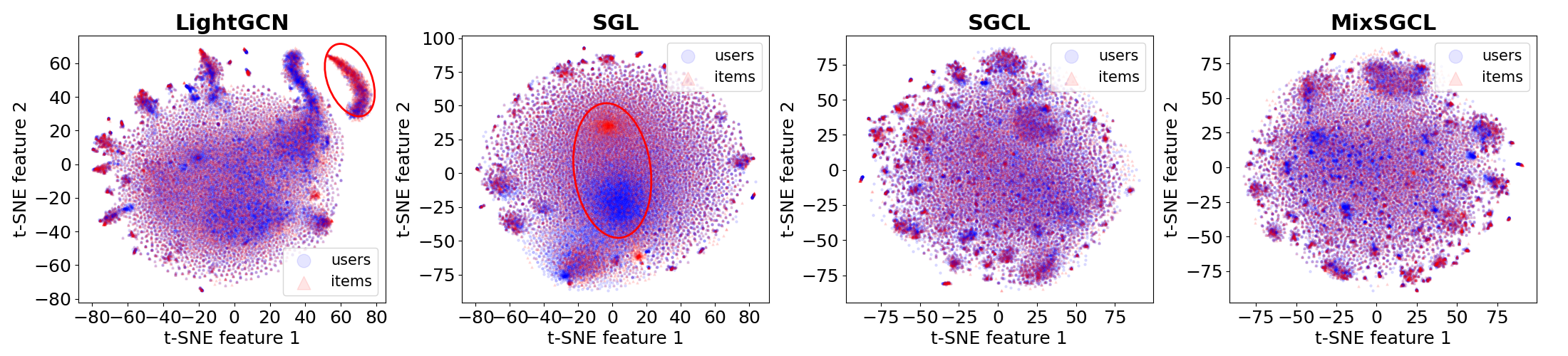}
        \caption{Visualization of the User-Item Embeddings on Toys-and-Games Dataset}
        \label{fig: visual_toys}
    \end{subfigure}
    \caption{The t-SNE visualization of the final embeddings of graph-based recommender systems in Beauty and Toys-and-Games datasets. The red circle highlights the most obvious distribution shift between the users and items embedding spaces.}
    \label{fig:graphs}
\end{figure*}

\subsection{Hyperparameter Analysis}

\subsubsection{\textbf{Impact of the Temperature $\tau$}}
\begin{figure}[htpb]
    \centering
    \includegraphics[width=1.0\linewidth]{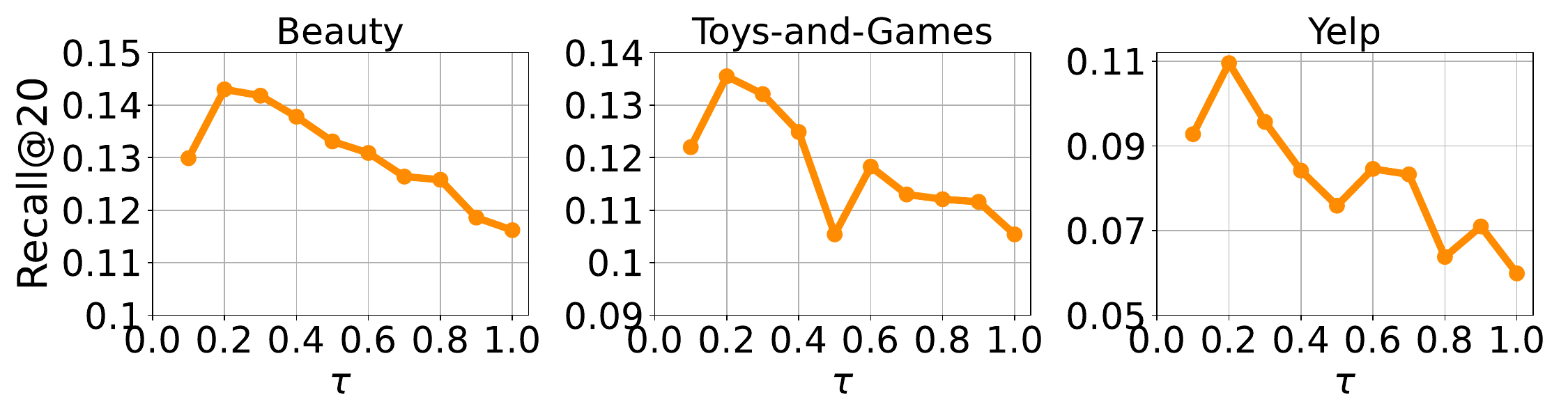}
    \caption{Impact of the temperature $\tau$}
    \label{fig: tem}
\end{figure}

Here, we showcase the influence of the temperature $\tau$ on the performance of our MixSGCL across three datasets in figure \ref{fig: tem}. On three datasets, when the temperature $\tau$ is tuned to 0.2, our model reaches the highest score of Recall. Then, the performance gradually declines as $\tau$ grows. Lower value $\tau$ increases the sensitivity of distinguishing negative pairs in the training batch, which is advantageous for the recommendation performance.

\subsubsection{\textbf{Impact of the Times of Augmentation $N_{mix}$}}
\begin{figure}[htpb]
    \centering
    \includegraphics[width=1.0\linewidth]{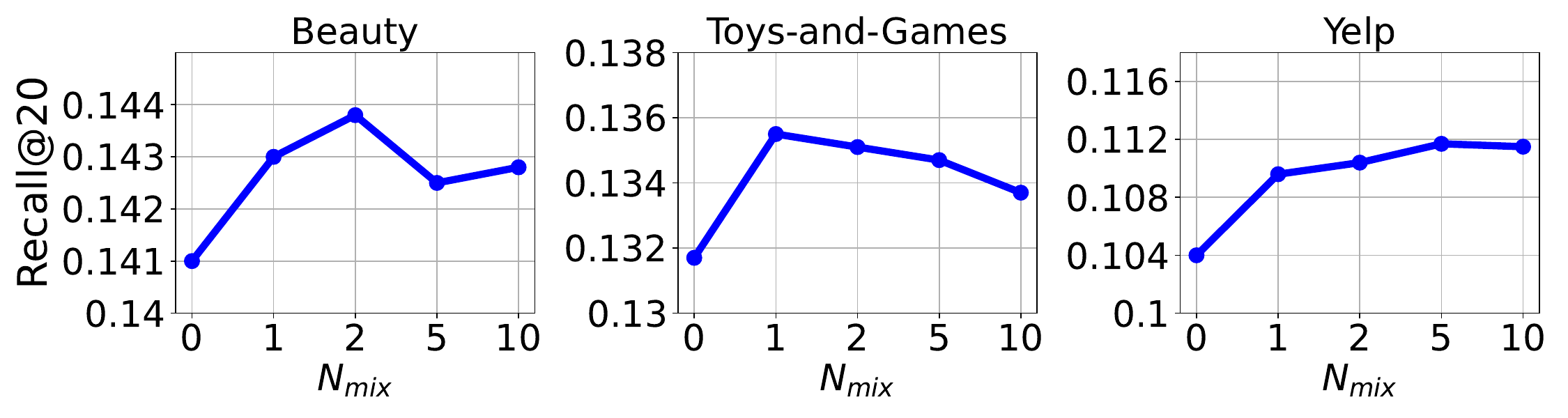}
    \caption{Impact of the number of augmentations $N_{mix}$}
    \label{fig: n_mix}
\end{figure}

To study how the number of times of augmentations influences the model training, we conduct the experiments of varying the $N_{mix}$ from 0 to 10. The results are reported in Figure~\ref{fig: n_mix}, where larger numbers of augmentation do not bring significant change in the ranking performance. Thus, we set $N_{mix}$ as 1 in all the other experiments considering the time efficiency.  

\section{Related Work}
\subsection{Graph Contrastive Learning for RecSys}

The data sparsity issue significantly limits the potential performance of recommendation models \cite{yin2020overcoming}. Recently, contrastive learning has gained considerable attention across various fields due to its proficiency in handling massive amounts of unlabeled data \cite{chen2020simple, tian2020makes, you2020graph, peng2020graph, hassani2020contrastive}. The essence of self-supervised learning lies in deriving valuable knowledge from large quantities of unlabeled data through meticulously crafted self-supervised tasks. Building upon this principle, a number of self-supervised recommendation models \cite{wu2021self,lin2022improving,yu2022graph,cai2023lightgcl} have been developed.
To address the issue of data sparsity, NCL\cite{lin2022improving} incorporates potential neighbors into contrastive pairs, by considering both the graph structure and semantic space. Different than NCL, the dropout-based methods are employed by SGL\cite{wu2021self} for randomized graph augmentation to obtain two contrastive views. The optimization includes node-level contrast using the InfoNCE loss, along with joint optimization using the BPR loss for recommendation. SimGCL\cite{yu2022graph} first investigates the necessity of complex graph augmentation and then devises noise-based augmentation techniques that perturb the representations of users and items during training, contributing to learning more robust features and preventing overfitting. However, the use of random perturbation in graph augmentation may result in the loss of valuable structural information. To alleviate this issue, LightGCL \cite{cai2023lightgcl} has been proposed to utilize the singular value decomposition (SVD) \cite{rajwade2012image} as a light and efficient tool to augment user-item interaction graph structures. Nevertheless, such augmentations still contain randomness in the randomized SVD algorithms, and some significant structure information can be dropped during the selection of low-rank singular values.

\subsection{Mixup}
Data augmentation is an effective method to enhance the generalization ability to unseen data \cite{shorten2019survey}, particularly in scenarios with noisy and scarce training datasets \cite{MAHARANA202291}. 
Mixup was first introduced for vision, speech, and tabular data \cite{zhang2017mixup} augmentation. The strategy of mixup is straightforward, which generates new data samples through mixing up the representation of two data samples and, similarly, their labels. Later works that adapt the idea of mixup to graph data include G-transplant \cite{park2022graph}, which selects subgraphs based on structural information and transplants subgraphs to target graphs; G-mixup \cite{pmlr-v162-han22c} which interpolates graphons to generate samples for classification; 
mixup for node and graph classification \cite{mixupcls} which designs a two-branch GCN layer to mix up the representation for nodes in a subgraph;
S-Mixup \cite{pmlr-v202-ling23a} employs soft alignment to compute the node correspondences and match nodes across graphs for mixup. In the realm of RecSys, MixGCF \cite{mixgcf} addresses the limitation of directly sampling negative data samples from the graph and presents their mixup strategy for hard negative data generation.
DINS~\cite{wu2023dimension} dimension-independently mixup the positive items and negative sampled ones to create hard negative sampling for model training. Therefore, neither methods \cite{mixgcf,wu2023dimension} are beneficial in self-supervised learning as they only aim to generate hard negative item samples in the BPR loss.
{Different from previous methods, we propose the novel node-level and edge-level intra-class mixup on positive data. It augments supervision signals directly from ground-truth interactions, which greatly alleviates the data sparsity issue.}

\section{Conclusion}

In this paper, we revisit the existing graph contrastive learning methodologies in the recommendation and identify two critical challenges, i.e., \textit{Inconsistent Gradients} and \textit{Unsupervised Augmentation}. Towards these issues, MixSGCL innovatively integrates the training of recommendation and contrastive losses using a supervised contrastive learning loss. This unified approach effectively resolves the problem of inconsistent optimization directions and suboptimal performance, typically seen in decoupled multi-task learning frameworks. Furthermore, by incorporating Node-wise and Edge-wise Mixup techniques, MixSGCL adeptly adopts a supervision-guided augmentation and tackles the prevalent data sparsity problem in recommender systems, extracting richer, supervised collaborative filtering signals from user-item interactions. The empirical validation of MixSGCL, conducted through extensive experiments on three real-world datasets, underscores its superior performance. Notably, MixSGCL outshines state-of-the-art methods, achieving faster convergence speed, and excelling in both accuracy and efficiency. This paper's findings highlight the potential of coupled design in supervised graph contrastive learning, paving the way for more efficient and effective recommender systems.

\bibliographystyle{ACM-Reference-Format}
\bibliography{sample-base}

\appendix


\end{document}